# Single-step modified electrodes for vitamin C monitoring in sweat

Bergoi Ibarlucea[1,2]*, Arnau Pérez Roig[1], Dmitry Belyaev[1], Larysa Baraban[1,2]*, Gianaurelio Cuniberti[1,2]

[1]Institute for Materials Science and Max Bergmann Center for Biomaterials, Technische Universität Dresden, Dresden (Germany)

[2]Center for advancing electronics Dresden (cfaed), Technische Universität Dresden, Dresden (Germany)

*Bergoi Ibarlucea: bergoi.ibarlucea@tu-dresden.de; Larysa Baraban: larysa.baraban@tu-dresden.de

**Abstract**

We demonstrate a flexible sensor for ascorbic acid detection in sweat based on single-step modified gold microelectrodes. The modification consists on the electrodeposition of alginate membrane with trapped CuO nanoparticles on top of the electrodes. The electrodes are fabricated at a thin polyimide support and the soft nature of the membrane can withstand mechanical stress far beyond the requirements for skin monitoring. We further show the efficient detection of ascorbic acid at the micromolar levels in both, a neutral buffer and acidic artificial sweat, at ultra-low applied potential (-5 mV). The effect of possible interfering species present in sweat is minimized, with no observable cross-reaction, thus maintaining a high degree of selectivity despite absence of enzymes in the fabrication scheme. This sensor is envisioned as a promising component of a wearable device for e.g. non-invasive monitoring of micronutrient loss through sweat.

**Introduction**

Ascorbic acid (vitamin C) is a long-time known important molecule, essential for correct metabolic functioning of the body (Carr and Frei, 1999). It participates in important processes such as synthesis and modifications of collagen (the main structural protein of hour body) (McLennan et al., 1988) or the reduction of nitrous acids in stomach preventing the appearance of carcinogenic compounds (Waring *et al.*, 1996). Like other water-soluble vitamins, ascorbic acid cannot be stored in the body for long time (Ganjali et al., 2017); it can be found in body fluids, .*e.g.* sweat (Lee *et al.*, 2017). Consequently, heat-exposed activities (*e.g.* in steel factories) or sports result in nutrient loss through sweating, including that of ascorbic acid (Tang *et al.*, 2016). Therefore, monitoring of micronutrients in sweat is convenient for the on-time response by dietary intake without delay in case of prolonged heat-exposed activities.

Thus, ascorbic acid is an important target for quantitative sensing and monitoring in the biomedical field, relevant also for the food and beverage industry, *e.g.* for quality control. Typically it is measured by high performance liquid chromatography (Nováková *et al.*, 2008), enzyme-based colorimetric multiwell kits (Chang *et al.*, 2019), or capillary electrophoresis (Tortajada-Genaro, 2012), with limits of detection and linear ranges oscillating between nanomolar and micromolar concentrations. These methods are remote laboratory techniques, with no possibility for on-site measurements.

Taking into account modern day lifestyle and needs, the sensor design principles should evolve correspondingly (Kamiŝalić *et al.*, 2018). In an increasing population or in special professional groups like athletes or heat-exposed workers, a miniaturized electrical device with low power requirements and low cost but retaining high sensitivity, integrated in the garment or the skin, would fulfill the requirements of a continuous monitoring for preventive healthcare application. In this context, micro- and nanofabrication techniques play an important role by helping to perform detection processes in a small and low-cost chip (Ibarlucea *et al.*, 2016) with the potential to include flexible characteristics to withstand mechanical stress produced by the movement of the end user (Gao *et al.*, 2016; Karnaushenko *et al.*, 2015; Miyamoto *et al.*, 2017; Zhang *et al.*, 2019).

Most of the recently reported biosensors for ascorbic acid are electrochemical, either enzymatic or nonenzymatic, as it will be described in the following lines. The first group makes use of the natural selectivity of enzymes to discriminate between analyte and possible interferents, being ascorbate oxidase the one used for the present case (Csiffáry *et al.*, 2016). This enzyme catalyzes the oxidation of ascorbic acid to dehydroascorbic acid by accepting electrons on its copper center, causing the electron movement that can be measured by the transducer. On the other side, nonenzymatic sensors make use of new materials to overcome the limitations of biological receptors in terms of cost and stability (Gnana Kumar *et al.*, 2017). Several attempts have pursued this approach for ascorbic acid detection, for example using carbon-supported PdNi nanoparticles (NPs) on glassy carbon electrodes *(*Zhang *et al.*, 2013), carbon nanoplatelets derived from ground cherry husks (Li *et al.*, 2017) or silver NPs grafted graphene/polyaniline nanocomposites (Salahandish *et al.*, 2019). Metal oxide NPs provide a high surface area and good electron transport kinetics *(George et al.*, 2018), which also make them a good candidate for the development of nonenzymatic sensors. Ascorbic acid is known to be a reducing agent for nanostructured copper, being used for NPs preparation (Cheng *et al.*, 2006). This property can be exploited for its detection, as demonstrated using CuO (copper oxide) nanowires synthesized on Cu foils by a thermal oxidation process at 350 °C for 100 min (You et al., 2019), or with 3D graphene/CuO nanoflowers fabricated by copper electrodeposition on a graphene structure grown by chemical vapor deposition (CVD) on a nickel foam that was later etched (Ma et al., 2014). Ideally, the cost and duration of the fabrication process should be minimized by avoiding of complex equipment and energy consuming steps and the sensor should operate at a low voltage and at room temperature. However, the aforementioned examples make use of either high fabrication temperatures, complex equipment (CVD), commercially unavailable NPs or nanostructures that need to be synthesized, or result in a sensor that measures at a high potential with a higher probability of oxidizing other molecules.

Here, we report on an ascorbic acid sensor based on CuO NPs, fabricated via electrodeposition of an alginate membrane on gold electrodes evaported on a light-weight polyimide support. The alginate membrane, which is reported to be biocompatible and permeable to the diffusion of the electrochemical substrates but impermeable to cells or large particles, has previously been used to trap enzymes for glucose and lactate detection (Márquez *et al.*, 2019, 2017) and carbon nanotubes to measure microbial activity (Mottet *et al.*, 2018). Here, we trap CuO NPs to make use of their natural interaction with ascorbic acid previously mentioned. The membrane is formed in few minutes in one step and its size can by controlled by the electrodeposition rate *(Cheng et al.*, 2011). Moreover, it can be easily removed by using a calcium chelating buffer like phosphate buffered saline (PBS), offering reusability of the electrodes to immobilize a new membrane with possibly different catalysts for additional measurements (Márquez et al., 2017). The use of simple gold electrodes on polyimide makes it compatible with a variety of techniques like inkjet printing or photolithography and with the future application as a flexible, wearable device. Finally, the detection of ascorbic acid is done at the micromolar range as it is found in sweat (Kilic *et al.*, 2017; Tang *et al.*, 2016), and in artificial perspiration solution without interference of any of the other present biomolecules.

## Materials and methods

*Electrode fabrication and modification*

Gold electrodes were first fabricated on a 100 µm thick polyimide film (Dupont, USA) by a standard photolithography process followed by 100 nm gold thermal evaporation on 3 nm chromium adhesion layer. The chip (Figure 1a) consisted of three electrodes, using the smallest one (0.3 mm$^2$) as working electrode so that the kinetics at the counter electrode do not limit those at the working one.

The alginate membrane precursor solution was composed by 1 wt% alginic acid sodium salt and 0.5 wt% calcium carbonate. 16 mg/mL CuO nanopowder (<50 nm) was added to the mix as catalyst for the ascorbic acid detection. 20 µL of the mix was drop casted on the electrodes and a galvanostatic deposition was carried out using a PalmSens4 potentiostat – galvanostat at a current density of 1.7 A/m$^2$ for 140 s. The same system was used for all next cyclic voltammetries and chronoamperometries. The anodic current produces a water electrolysis generating protons that react with the calcium carbonate, which tends to buffer the solution by consuming the protons and forming $Ca^{2+}$ ions and $CO_2$ (Cheng *et al.*, 2011). The calcium ions interact with the alginate polymers by cross-linking them and creating the membrane, where the CuO NPs get trapped (Figure 1a-i). The process resulted in a flexible chip with a membrane on the working electrode (Figure 1a-ii). The drop with the precursor mix could be recovered and used for the deposition on further electrodes. Subsequently, the formed membrane was rinsed in 145 mM NaCl and 10 mM imidazole buffer supplemented with 145 mM KCl to mimic physiological conditions and $CaCl_2$ to prevent $Ca^{2+}$ ion loss and therefore membrane dissolution.

*Alginate membrane characterization*

The membrane was characterized by optical microscopy, scanning electron microscopy (SEM) and cyclic voltammetry.

Larger current densities than 1.7 A/m$^2$ led to electrode delamination, while smaller ones produced weak and unstable membranes that are easily removed. Shorter electrodeposition times (60 s) resulted in soft membranes from which the CuO NPs were easily washed away (Figure S1, SM), while above 140 s the membrane was adsorbed to the electrode in a stable manner during the rinsing steps and the NPs covered the whole working electrode. The resulting membrane after 140 s had a thickness of *ca.* 10 µm as measured via digital microscopy (Figure 1a, iv) with a toluidine blue O stained membrane to confer a more homogeneous and non-particulated contrast.

The membranes were prepared for SEM observation following a protocol for hydrogel fixation reported elsewhere (Muri *et al.*, 2018) but replacing the used buffer for the imidazole based one to avoid membrane dissolution. First hydrogels with and without CuO NPs were fixated in 2.5% glutaraldehyde in imidazole buffer for 2 h at room temperature. They were then left at 4 °C for 12 h. After washing twice for 5 min. in imidazole buffer, the samples were dehydrated in solutions of increasing ethanol concentration (10, 25, 50, 70, 90, and 100%) for 5 min. for each concentration. Later, the alginate hydrogel was incubated for 20 min in 50% HMDS diluted in ethanol followed by an additional incubation in a new solution with same characteristics. Finally, the double incubation was repeated in two 100% HMDS solutions. Membranes were left to dry overnight in a desiccator and coated with a 20 nm Au layer by thermal evaporation.

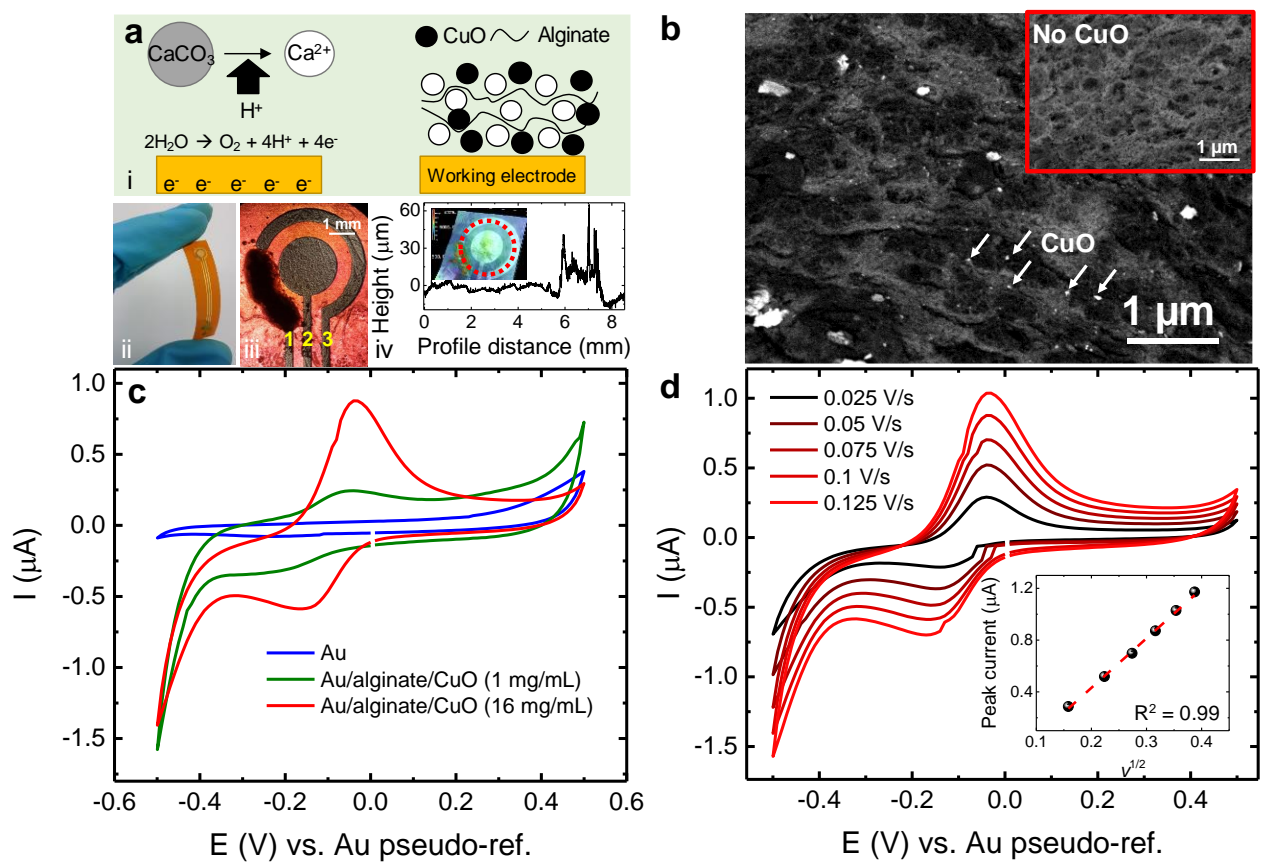

**Figure 1.** (a) Alginate electrodeposition with CuO nanopowder. (i) Scheme of electrodeposition mechanism. The flexible chip is shown in (ii), with the membrane formed after 140 s electrodeposition time. Yellow numbers indicate electrode function: (1) working, (2) counter, (3) pseudo-reference. Film thickness (10 µm) is shown in (iv), measured along the red dotted circle in inset. (b) SEM imaging of alginate membrane with CuO NPs. Back-scattered electrons provide the contrast, with the white spots belonging to the CuO (arrows indicate some examples). Inset shows the membrane without CuO, where no white spots can be seen. (c) Cyclic voltammogram on bare electrodes (blue line) and modified electrodes (red and green), showing the appearance of the redox pair belonging to CuO. (d) Scan rate test and calibration (inset) showing linear relation with square root of the scan rate.

*Bending test*

The flexibility of the sensor owing to the mechanical properties of the thin polyimide foil and the soft nature of the alginate was tested by performing cyclic voltammetries before and after controlled bending events. For this, the chip with the electrodeposited membrane was placed on a cylinder of 5 mm radius and bent 1000 times in 10 cycles of 100 bending events each. Cyclic voltammetry was performed between each cycle to observe variations in the signal.

*Ascorbic acid detection*

All measurements in buffer were done in the aforementioned imidazole system as non-chelating buffer, adjusted to pH 7.4. First, cyclic voltammetry at 0.1 V/s was used to observe the effect on the signal produced by the effect of ascorbic acid. Then, the optimal working potential was chosen for amperometric measurements with different ascorbic acid concentrations at constant pH, to calibrate the sensor and determine the concentration range that could be measured. Next, we proceeded to measure in artificial sweat samples. For this, an artificial perspiration solution was prepared based on the work by Kilic et al. (Kilic *et al.*, 2017), which was composed of 20 g/L NaCl, 17.5 g/L NH4Cl, 5 g/L acetic acid, 15 g/L lactic acid, 0.17 mM glucose, 59 uM uric acid, 0.18 mM pyruvic acid, 0.37 mM glutamic acid, and 10 mM urea. As sweat is a mildly acidic fluid (Coyle *et al.*, 2009; Curto *et al.*, 2012), the pH was adjusted to pH 5.5. Prior to the measurements directly in this solution, the possible interference by the presence of the organic molecules in the recipe was analyzed. Finally, the artificial perspiration solution was spiked with ascorbic acid and measured in the same way.

**Results and discussion**

*Sensor characterization*

SEM imaging of backscattered electrons showed the contrast between the membrane material and the trapped CuO (Figure 1b). The porous membrane appears dark, with white spots belonging to the CuO nanoparticles and aggregates of heterogeneous sizes. In absence of CuO NPs (inset to Figure 1b) no white spots were found.

A cyclic voltammetry of the resulting CuO-containing membrane showed the peaks of the redox pair corresponding to the presence of the CuO NPs (Figure 1c). The scan rate test between 25 mV/s and 125 mV/s showed a linear increase ($R^2 = 0.99$) of the peak amplitude with the square root of the scan rate and with no peak shift (Figure 1d), indicating a reversible and diffusion controlled system.

*Bending test*

The measured cyclic voltammograms (Figure 2a) did not show any significant horizontal shift of the redox peak position, which is the effect that allows measuring ascorbic acid in this work. The amplitude of the anodic peak (Figure 2b) suffered a slight decrease of 1 µA during the first 300 bending events and remained constant for the next 700 bends. The amplitude decrease for the cathodic peak instead, decreased less than 0.5 µA, showing a high performance even after such high deformation. It must be noted, that 5 mm is far beyond the required radius for skin biomonitoring (Schwartz *et al.*, 2013).

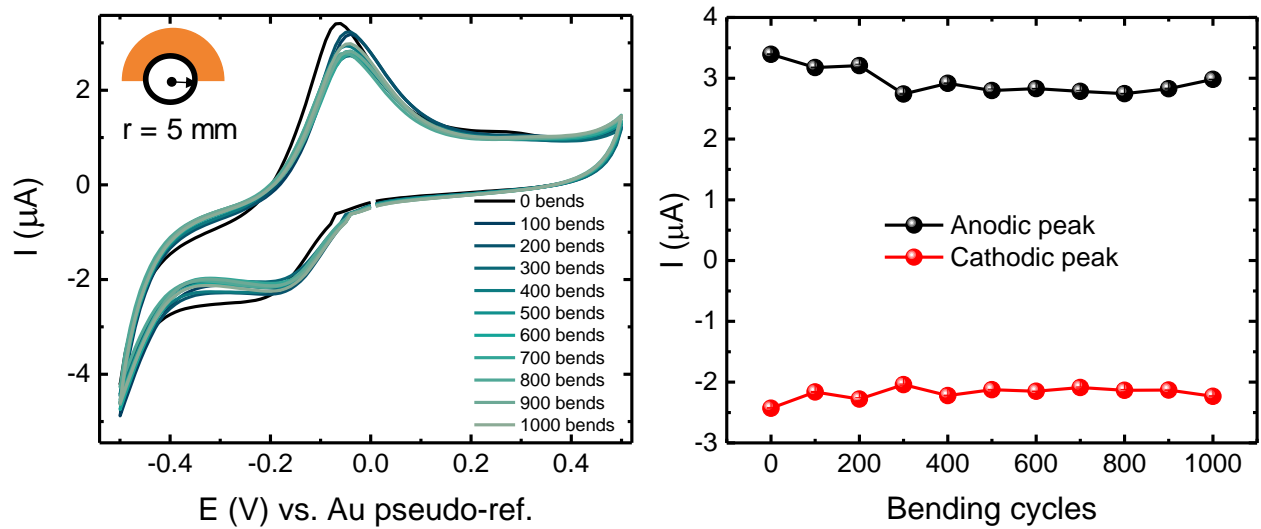

**Figure 2.** Results of the bending test at 5 mm radius. (a) Cyclic voltammograms and (b) redox peak position before and after each bending cycle. No peak shift was observed (basis for the ascorbic acid detection). After an amplitude decrease of less than 1 µA during the first 300 bends, it remained constant for the next 700 events. Inset in (a) shows the principle of the test. The orange film represents the chip on polyimide material, bent onto a 5 mm radius.

*Ascorbic acid detection in buffer system*

First, a comparison of the cyclic voltammogram was done in absence and presence of ascorbic acid (37 µM). Ascorbate produced a pronounced shift of the redox peaks to a positive potential direction (Figure 3a), indicating a migration of the CuO through different oxidation states. The fact that the current measured at zero volts in the backwards sweep of the cyclic voltammetry belongs solely to the capacitive current, makes it possible to set this potential in a chronoamperometry and measure the shift of the reducing peak as an increasing current as it shifts and approaches the working potential. Therefore, a chronoamperometry was performed at nearly zero volts (-5 mV), minimizing the possibility to oxidize other possible interfering biomolecules in a real sample. The imidazole buffer containing ascorbic acid in a range between 0 and 125 µM was measured, being 1.8 µM the smallest detected concentration. The samples were drop casted on the sensor and the chronoamperometry was carried out for 60 s for each concentration, observing an increasing negative current together with the increase in analyte concentration (Figure 3b). The linear range was observed to be up to 30 µM (Figure 3c), allowing to perform measurements in the range where ascorbic acid is found in sweat (10 – 30 µM) (Kilic *et al.*, 2017; Tang *et al.*, 2016).

A reproducibility test with membranes prepared with 6 mg/mL CuO showed on one hand that the measurable range was constant (Figure S2a, Supplementary Material, SM), and on the other hand that such concentration change does not alter the results, being CuO in molar excess. However, further reduction to 1 mg/mL (voltammogram at Figure 1c, green line) led to non-reproducible results (variable measurable range). We hypothesize that this is caused by the different CuO aggregate formation on each new membrane, which is also shown by the fact that current levels varied in samples with higher CuO concentration, although the measurable range was the same and the current levels are maintained in the same order of magnitude. With smaller CuO concentrations (1 mg/mL), the error coming from the uncontrolled aggregate formation is increased.

The quantification is also possible by measurements of either cathodic or anodic peaks. Figure S2b, SM, shows an example of the shift between 1 and 310 µM, with the calibration of the cathodic peak shift for 4 independent sensors in Figure S2c, SM, which speaks for the good reproducibility as well.

With measurements of real samples in mind, we performed an interference test by repeating the chronoamperometry in presence of several biomolecules typically found in sweat: lactic acid, uric acid, pyruvic acid, glutamic acid, glucose, and urea. Buffer solution was drop casted on the sensor and the measurement was started. After 50 s, additional buffer was drop casted with each of the biomolecules, reaching the final concentration found in the artificial perspiration solution to be used later on, including various acids (lactic, uric, pyruvic, and glutamic). The results are shown in Figure 3d. No response was found for any other than ascorbic acid, being this the smallest concentrated one (10 µM). Highly concentrated species at 10 mM (lactic acid, glucose, urea) also did not react. The response for the ascorbic acid was observed instantaneously, showing the fast response of the sensor.

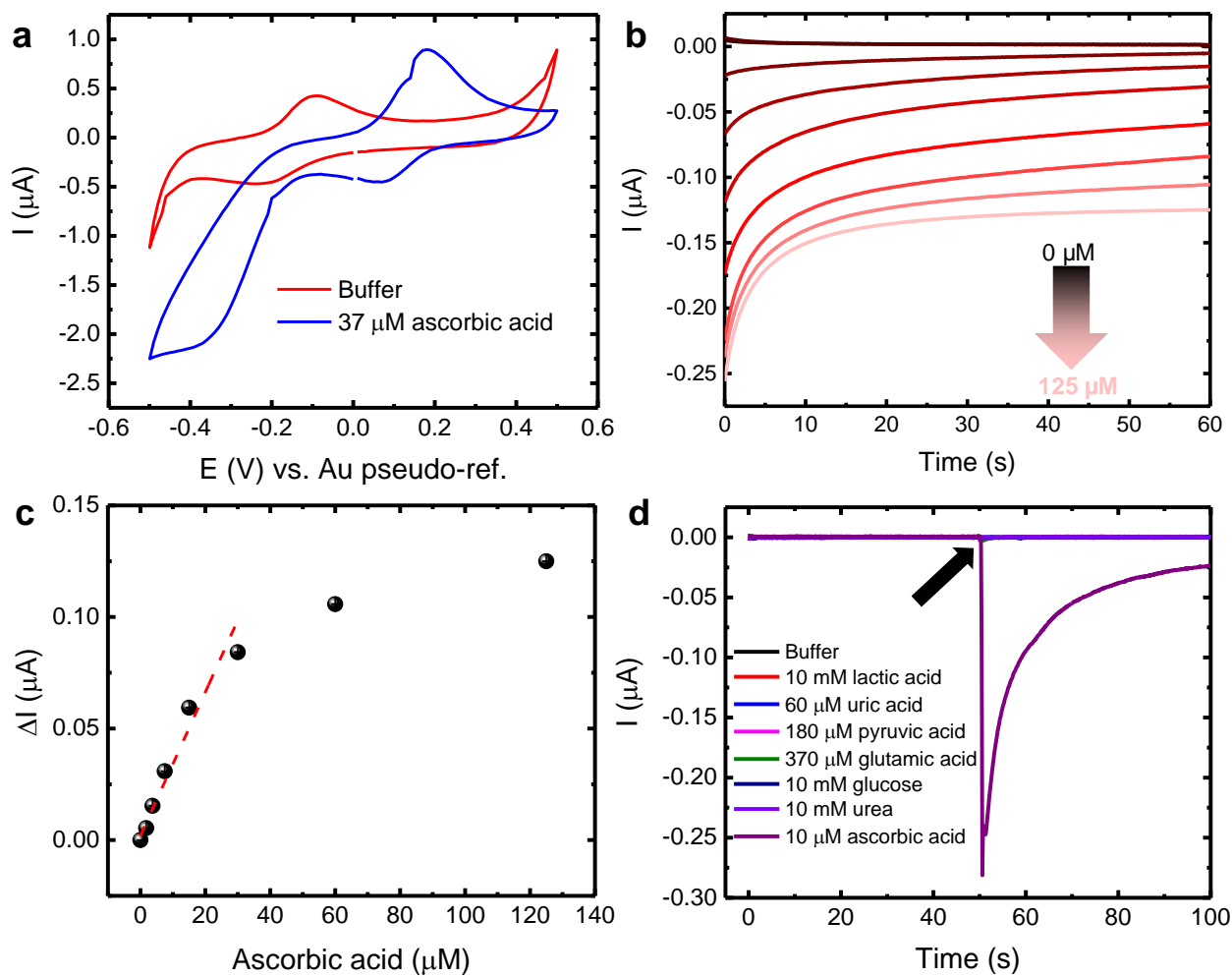

**Figure 3.** Ascorbic acid detection and analysis of the effect of the interferents. (a) Effect of the ascorbic acid on the cyclic voltammogram. The peaks of the CuO redox pairs are shifted towards positive potential. (b) Chronoamperometry at -5 mV in a range between 0 and 125 µM (smallest detected concentration = 1.8 µM). (c) Calibration of the chronoamperometry. (d) Interference test showing no response with other than ascorbic acid. The arrow indicates the interferent injection moment.

*Ascorbic acid detection in artificial samples*

Artificial perspiration solutions were prepared with increasing ascorbic acid concentrations and the chronoamperometry was performed at -5 mV. Compared to measurements in imidazole, here the measurable range was extended to higher concentrations, with saturation after ca. 300 µM (Figure 4a,b), which would allow detecting the levels found also in blood. The current levels were also increased, which is likely coming from the highly complex nature of the solution. The cyclic voltammograms shown in Figure S3a, SM, support this idea, where the redox peaks and the capacitive current where found to be 4-fold and 2-fold stronger than in imidazole respectively. In spite of that, the levels found in sweat could also be detected, as seen in the real-time measurement of 10 μM ascorbic acid (Figure 4c). By performing the complete cyclic voltammetry for each ascorbate concentration, the quantification should also possible by measuring the peak shift (Figure S3b, SM). Furthermore, one hour incubation in sweat did not produce any degradation of the signal (Figure S4, SM), demonstrating the stability of the system.

Finally, taking in account the complexity of sweat delivery on a wearable device, we propose the combination of the sensor with a lateral flow approach (Koczula and Gallotta, 2016), where a hydrophilic paper allows the gradual passing of a liquid sample. As preliminary test, a rectangular piece of filter paper (413-type filter paper, VWR, Germany) was directly placed on top of the sensor and 10 μL samples of artificial sweat with 100 μM ascorbic acid were dropped on one side the paper (inset to Figure 4d), which absorbed and redirected it towards the sensor, forming clear signal peaks while no signal was observed for the samples without ascorbic acid (Figure 4d).

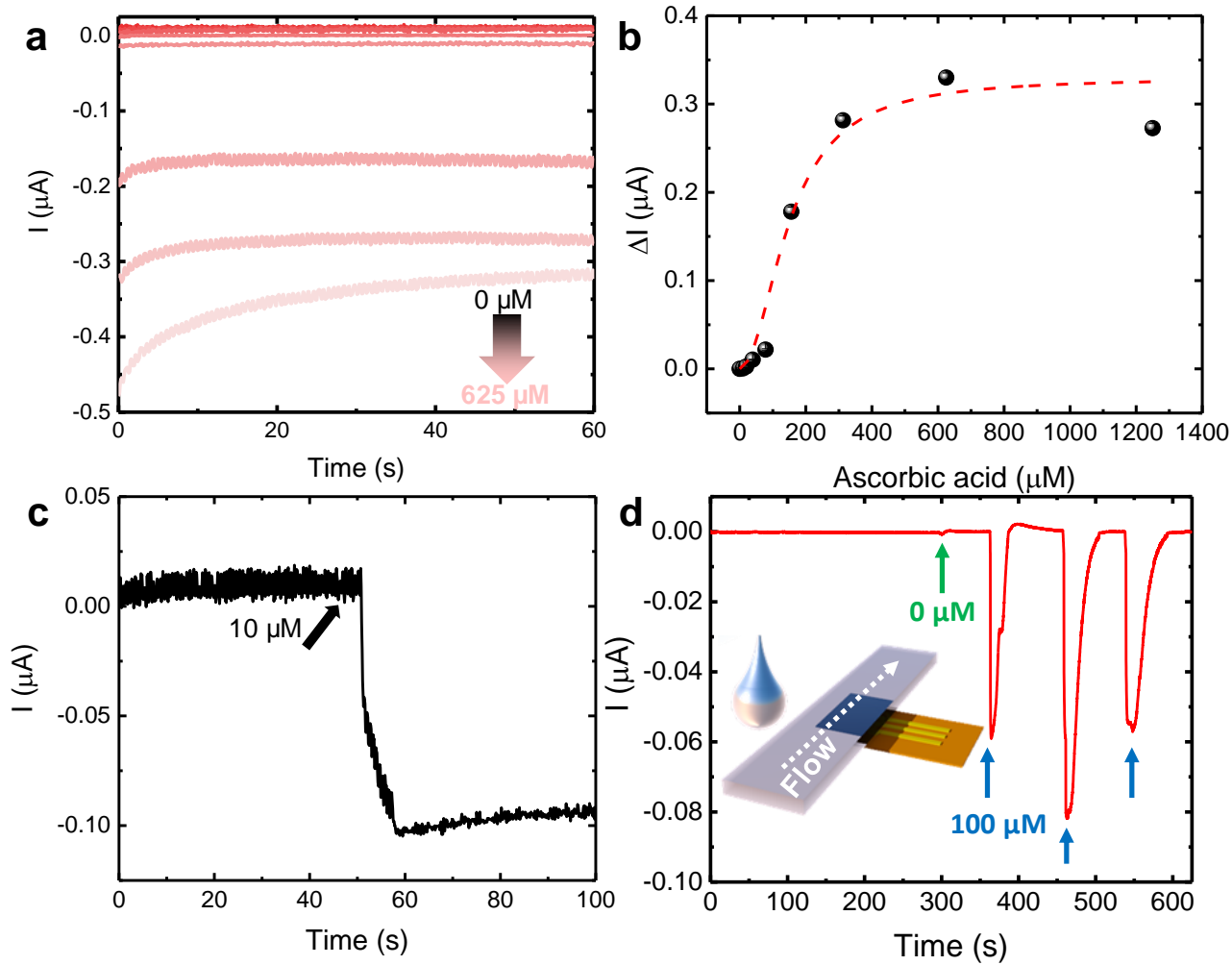

**Figure 4.** Measurements in artificial perspiration solution. (a) Chronoamperometry and (b) calibration. (c) Real-time measurement of 10 μM ascorbic acid (AA) by direct drop casting. (d)

Real-time measurement of sweat with and without ascorbic acid by sample delivery through filter paper.

Despite the possibility of measuring in the levels and physiological conditions in serum or blood, incubations in fetal bovine serum led to gradual degradation and disappearing of the redox peaks. A previously reported work with CuO entrapped in alginate together with glucose oxidase for glucose sensing however showed successful glucose detection without CuO degradation and ascorbic acid interference, which denotes that the presence of the enzyme might play a role in NP protection. In our case, a direct interaction between CuO and glucose is discarded in the degradation process, as demonstrated in the interference test and the measurements in artificial perspiration solution containing solution, and considering that such interaction requires alkaline conditions (Romeo *et al.*, 2018).

Other works exist with good sensitivity at micromolar concentrations and without enzymes, but require either working at a higher potential (Salahandish *et al.*, 2019) or a more difficult and longer fabrication process involving several steps (Pakapongpan *et al.*, 2014), high temperatures (higher cost) (Li *et al.*, 2017; You *et al.*, 2019), or complex/costly equipment (Ma *et al.*, 2014). Our proposed approach performs the measurements in both imidazole (neutral pH) and artificial perspiration solution (acidic) with a sensor response at nearly zero voltage and in the lower micromolar range, convenient for direct detection in human sweat.

**Conclusions**

We demonstrate a nonenzymatic sensor for ascorbic acid detection working at ultra-low potential and with sensitivity in the concentration levels found in sweat. The sensor is fabricated by a single-step electrodeposition process in only two minutes and using low-cost materials on a light-weight support. This approach is shown to be very effective in sweat, which is also a convenient fluid. On one hand, the measurement can be done in real patients with a less invasive method, and with the possibility to be implemented on a wearable device for real-time measurements. On the other hand, the acidity of sweat is a convenient property for the alginate membrane, which would dissolve in an alkaline solution. The measurements were possible in both neutral pH using the imidazole buffer and acidic artificial perspiration solution and show a detectable range similar to that found in real sweat, in nearly zero (-5 mV) ultra-low potential, without interference of many other biomolecules in the solution. Integration of other catalysts or enzymes for the detection of additional micronutrients and a further study of the sensor to analyze its mechanical properties, followed by the integration on a wearable electrical device with a miniaturized potentiostat would enable its implementation in sports or heat-exposed working environments of large duration, protecting the health of the person involved by a timely dietary ascorbic acid intake.

## Acknowledgements

This work was funded by the AiF project F009438. The authors thank the technical assistance of Dr. Petr Formanek (Leibniz Institute of Polymer Research Dresden e. V.) in the SEM imaging and Dr. Raphael Doineau (Leibniz Institute for Solid State and Materials Research) in the membrane thickness measurement.

Supplementary Material

# Single-step modified electrodes for vitamin C monitoring in sweat

Bergoi Ibarlucea[1,2]*, Arnau Pérez Roig[1], Dmitry Belyaev[1], Larysa Baraban[1,2]*, Gianaurelio Cuniberti[1,2]

[1]Institute for Materials Science and Max Bergmann Center for Biomaterials, Technische Universität Dresden, Dresden (Germany)

[2]Center for advancing electronics Dresden (Cfaed), Technische Universität Dresden, Dresden (Germany)

*Bergoi Ibarlucea: bergoi.ibarlucea@tu-dresden.de; Larysa Baraban: larysa.baraban@tu-dresden.de

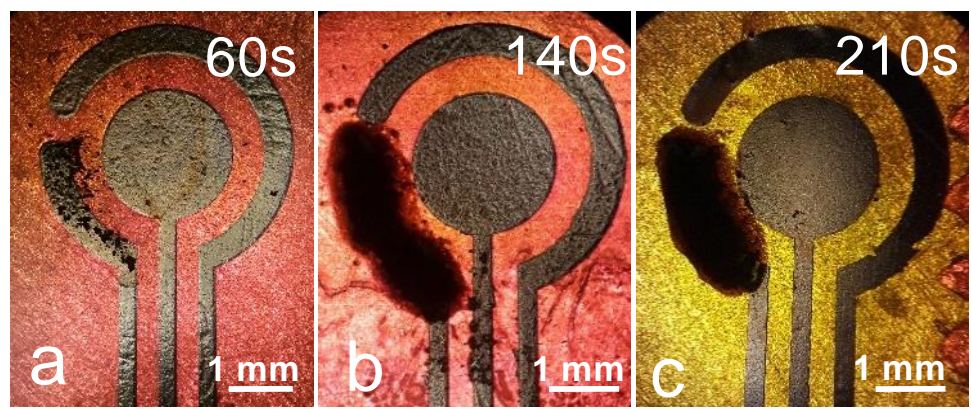

**Figure S1**. Membrane electrodeposition at (a) 60 s, (b) 140 s, and (c) 210 s.

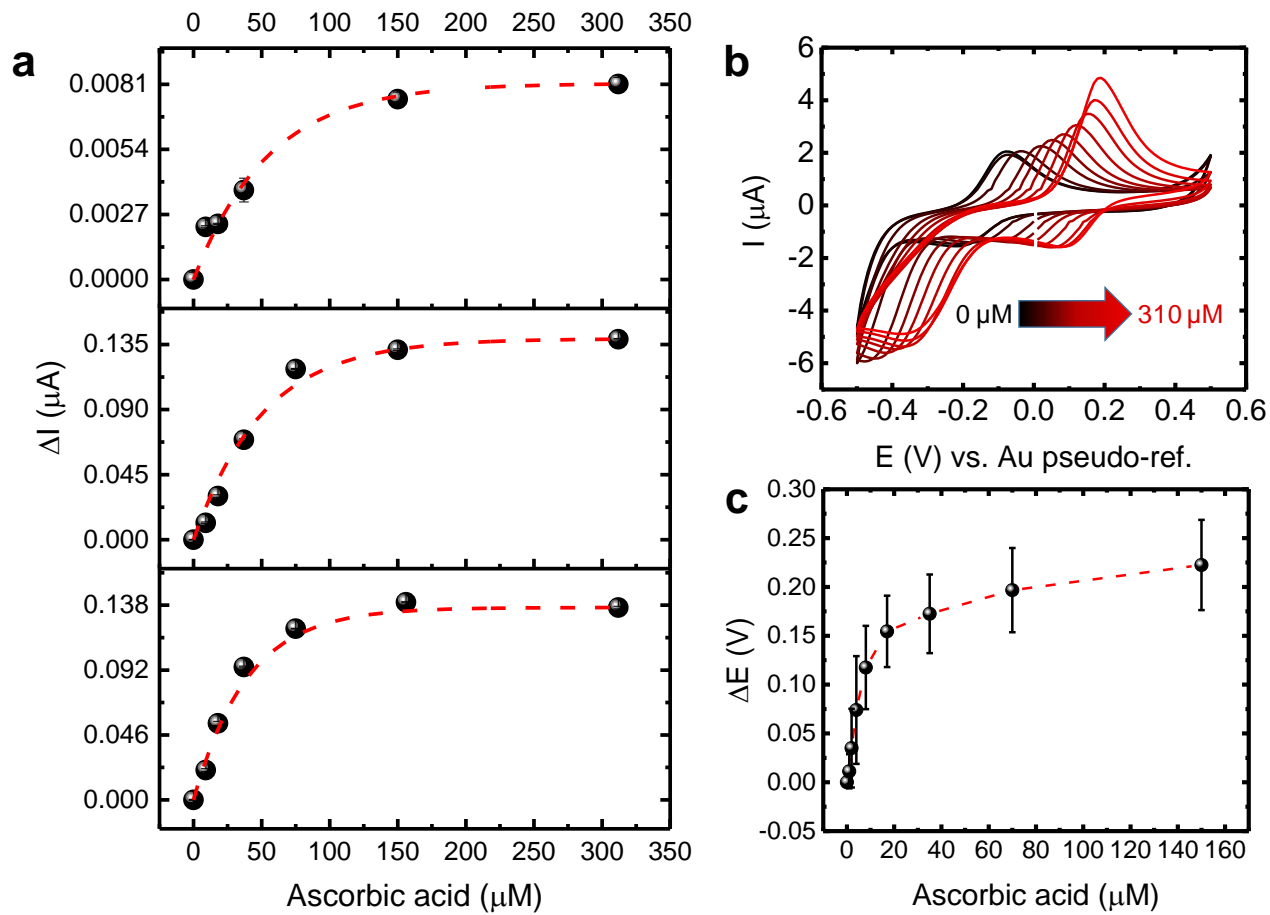

**Figure S2**. Reproducibility test. (a) Three additional sensors with the same detection range in buffer. (b) Cyclic voltammogram shift as example of one sensor. (c) Cathodic potential peak shift calibration for 4 independent sensors.

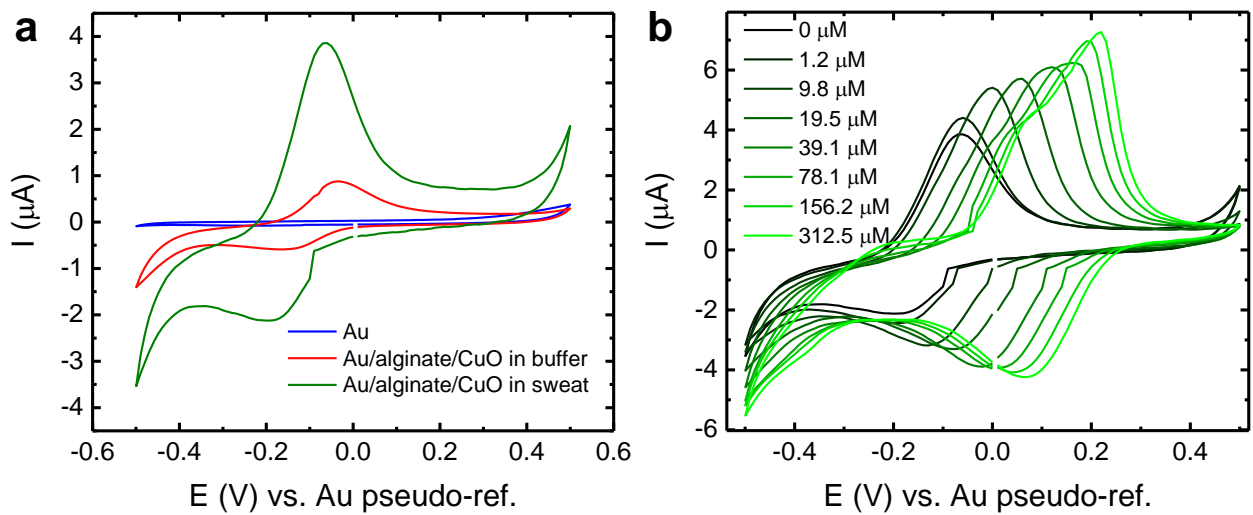

**Figure S3.** Behavior of the alginate membrane with CuO nanoparticles in artificial perspiration solution. (a) Cyclic voltammograms comparing bare electrodes, alginate and CuO modified electrodes in imidazole buffer, and modified electrodes in artificial sweat. (b) Cyclic voltammograms in artificial sweat with increasing ascorbic acid concentration.

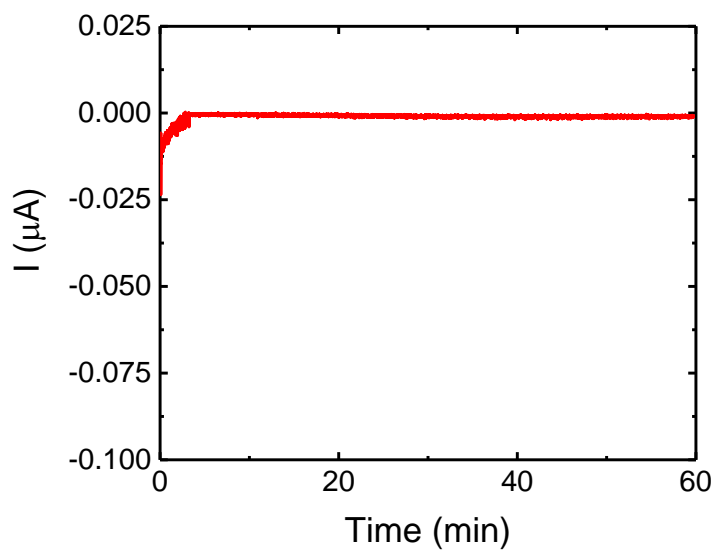

**Figure S4.** Stability test of the sensor by performing chronoamperometry in artificial sweat for one hour. No alteration of the signal is observed.